\documentclass[a4paper, 11pt, one column, aas_macros]{article}

\usepackage[english]{babel}
\usepackage[utf8x]{inputenc}
\usepackage[T1]{fontenc}
\usepackage{aas_macros}
\usepackage{multicol}

\usepackage[top=0cm, bottom=2.0cm, outer=2.5cm, inner=2.5cm, heightrounded,
marginparwidth=1cm, marginparsep=1cm, margin=1.5cm]{geometry}

\usepackage{caption}
\usepackage{graphicx} 
\graphicspath{ {./Images/} }
\usepackage[colorlinks=False]{hyperref} 
\usepackage{amsmath}  
\usepackage{amsfonts} %
\usepackage{amssymb}  %
\usepackage{multirow}
\usepackage{caption}
\usepackage{subcaption}
\usepackage{graphicx}
\usepackage{csquotes}
\usepackage{biblatex}

\bibliography{main.bib}

\title{Pupillary reactions depend on disgust sensitivity in conceptual pavlovian disgust conditioning}
\author{{\rm Lars OM Rothkegel}\\Universität Potsdam\\
\and {\rm Jakob Fink-Lamotte}\\Universität Potsdam}

\date{}
\begin{document}
\maketitle
\begin{abstract}

Exposure-based interventions rely on inhibitory learning, often studied through Pavlovian conditioning. While disgust conditioning is increasingly linked to psychiatric disorders, it has been less researched than fear conditioning. In this study, we applied a categorical Pavlovian disgust conditioning paradigm with two CS categories (animals and tools) and disgusting images as US (e.g., feces). During categorization, acquisition, and extinction phases, we measured eye movements and pupil responses in 44 participants.

Consistent with previous results, subjective disgust and US expectancy increased from categorization to acquisition for CS+, along with greater pupil dilation for CS+ than CS-. Higher disgust sensitivity was associated with more generalized and longer lasting disgust experiences, as well as higher expectancy for disgusting images. Pupil response during acquisition and extinction depended on disgust sensitivity: Participants with lower disgust sensitivity showed greater pupil dilation.

These findings suggest that individuals with high disgust sensitivity and prolonged expectancy may exhibit physiological differences from less sensitive individuals as early as the acquisition phase. This could inform contamination-based OCD treatments by integrating interventions which focus on attentional deployment or physiological reactions. To our knowledge, this is the first study using pupillometry and eye tracking in categorical disgust conditioning.

\end{abstract}
\section{Introduction}
Obsesive-compulsive-disorder (OCD), a mental health condition affecting approximately 2–3\% of the population \parencite{horwath2022epidemiology}, is characterized by intrusive, unwanted thoughts (i.e. obsessions) that drive repetitive, ritualistic behaviors (i.e. compulsions; \cite{american2013diagnostic}). Exposure and response prevention (ERP) therapy is the gold standard for treating OCD \parencite{hirschtritt2017obsessive}, but about 30\% of individuals show little to no improvement with ERP \parencite{abramowitz2003symptom}. One reason for this lack of succesful treatment might be, that exposure therapy is based on inhibitory learning of fearful associations \parencite{craske2014maximizing}. However, research of the last 20 years has identified disgust as an important aspect regarding the development and maintainance of OCD \parencite{mason2010looking,knowles2018disgust} and thus presumably an important part of succesful treatment \parencite{mason2012treating,ludvik2015effective,rast2023treatment}. Additionally, research has shown that disgust is more resistant to extinction than fear \parencite{mason2010looking,mitchell2024disgust}, which could explain why ERP in the way it is commonly applied does not work for a substantial number of OCD patients. One type of OCD specifically related to disgust is contamination-based OCD, also known as C-OCD. C-OCD, one of the most common types of OCD, involves persistent concerns about germs and compulsive washing or sanitizing behaviors \parencite{ruscio2010epidemiology}. The Covid-19 pandemic has even increased the importance of this debilitating disorder \parencite{guzick2021obsessive}. 


To illustrate, how disgust can play a role in the development and maintainance of C-OCD, imagine the following screnario: 
You walk to a bus station in the morning and unexpectedly encounter a puddle of vomit. Your immediate reaction might involve wrinkling your nose, turning away, or even experiencing a gag reflex. Your attention and thus your gaze will presumably first be directed to the puddle due to attentional capture \parencite{fink2022you,van2013disgust} and, following this initial reaction, be directed away to avoid further aversive feelings \parencite{armstrong2022ve}. Additionally, this encounter presumably will elicit notable physiological responses. Your parasympathetic nervous system may activate, increasing digestive system activity \parencite{de2011sympathetic}, and constricting pupil size \parencite{wang2021generalization}. Simultaneously, the sympathetic nervous system could increase heart rate and skin conductance levels \parencite{stark2005psychophysiological}. On a cognitive level, you might think something like \textit{urgh, this smells and I don't want to sit here, because I could get sick from breathing in these germs.} All these automatic responses to the stimulus underscore the evolutionary function of disgust as an adaptive response that protects individuals from harmful or contaminated stimuli  \parencite{tybur2013disgust}. 

While such an experience may result in the development of an aversive association with - for example - the location where it was encountered through classical conditioning \parencite{pavlov2010conditioned}, subsequent visits to the again-clean bus stop are likely to diminish this negative feeling as inhibitory learning overwrites the aversive association \parencite{craske2014maximizing}. However, coming back to how disgust and OCD might be related, disruptions in inhibitory learning — due to e.g. individual predispositions or unfavorable attentional/physiological responses — can lead to persistent disgust reactions \parencite{mitchell2024disgust}. One of these predispositions could be disgust sensitivty (DS),  i.e. the tendency to perceive disgust as particularly aversive \parencite{van2006disgust}. DS has been shown to be related to several OCD symptoms, with the clearest relationship between C-OCD symptoms and disgust sensitivity toward hygiene-related stimuli \parencite{tolin2006disgust}. Additionally, without any inhibitory learning and potential subsequent cognitive associations (i.e. \textit{If this bus stop if so disgusting, maybe other bus stops are equally disgusting}), the aversive feeling can generalize to other locations, such as other bus stops and train stations \parencite{dunsmoor2015fear}. 
Thus, one particular factor driving psychopathology in OCD might be the generalization of disgust associations to other stimuli. This has for example been shown for panic disorder and fear generalization \parencite{lissek2010overgeneralization}. Generalization learning involves the transfer of a conditioned response (CR) to stimuli resembling the original conditioned stimulus (CS; \cite{dunsmoor2015fear}). While most experiments on generalization learning have focused on perceptual similarities (e.g., sensory features such as variations in circle sizes; \cite{jasnow2017perspectives,wang2021generalization}) generalization can also occur at a categorical or cognitive level (going back to the thought of \textit{probably all other bus stops and trains stations are also dirty!}). Dunsmoor et al. (\citeyear{dunsmoor2015emotional}) demonstrated that emotional learning at a categorical level can retroactively enhance memory consolidation for an entire group of stimuli, incorporating non-reinforced stimuli into the mental association. This finding highlights how neutral and aversive stimuli can form associations without a clear perceptual link. This conceptual learning has recently been shown to influence disgust-associations \parencite{wang2024conceptual}. 

To shed new light to the question of why and for whom disgust associations are more resistant to inhibitory larning, the goal of this study was to investigate how cognitive, affective, phyisological and attentional responses during conceptual disgust conditioning (similar to Wang et al., \citeyear{wang2024conceptual}) are shaped by interindividual differences, in this case disgust sensitivity. Before describing the Experiment and corresponding results, we will give a short overview about the differential responses measures

\textbf{Cognitive response during disgust conditioning}
In their comprehensive review, Knowles et al. (\citeyear{knowles2019cognitive}) have provided a representation of the different types of cognitive biases associated with disgust and their relation to psychopathology. One bias they describe is the so called expectancy bias, i.e. the bias to exaggerate the likelihood of a negative event to occur \parencite{davey1995preparedness,foa1996cognitive}. Knowles et al. state that \textit{Indeed, negatively biased expectations of disgust-relevant outcomes (i.e., disease) may be a key component in the development of anxiety by motivating avoidance behavior \parencite{reiss1991expectancy}}. Expectancy for an aversive event to occur is the logical variable for a cognitive bias in a conditioning experiment, since we have direct control over the actual occurence of aversive events and how this affects cognitive evaluation. Armstrong et al. (\citeyear{armstrong2017pavlovian}) already demonstrated, that individuals high in disgust sensitivity show delayed extinction learning in a Pavlovian disgust conditoniong experiment and Wang et al. (\citeyear{wang2024conceptual}) have shown that this also applies in a conceptual disgust conditioning experiment regarding participants with low Vs. high contamination fears. We thus expect, that in our experiment high disgust sensitivity leads to delayed extinction learning represented by stronger ad longer lasting expectancy for a disgusting stimulus to occur. However, instead of splitting our sample in two groups we expect that disgust sensitivity influences US-Expectancy as a covariate in a linear mixed model in an extinction phase (H1).

\textbf{Affective response during disgust conditioning}
The subjective evaluation of how disgusting a stimulus is rated represents the affective component in our study. Participants with higher disgust sensitivity have been shown to rate disgust stimuli as more disgusting \parencite{haberkamp2017disgust} and rate stimuli, which predict disgusting stimuli (i.e. CS+), as more disgusting \parencite{wang2024conceptual}. We aimed at replicating previous results and transfering them to a conceptual disgust conditioning paradigm and thus hypothesized that heightened disgust sensitivity would lead to higher disgust rating of the CS stimuli. Again this will be measured with disgust sensitivity as a covariate within linear mixed models instead of a split sample (H2).

\textbf{Physiological response during disgust conditioning}
As stated in the introductury example, disgust evokes many physiological responses. One rather fast physiological response is the pupilary reaction \parencite{mathot2018pupillometry}. Besides light influences, emotional and cognitive aspects can shape pupillary responses \parencite{mathot2018pupillometry}. Since pupillary constriction is connected to the parasympathetic nervous system and dilation to the sympathetic nervous system, the pupil is an interesting object of investigation regarding disgust. While a previous study did not find a connection between pupil size and image type when comparing disgusting and fearful images, they found that disgust sensitivity was negatively correlated with pupil size \parencite{schienle2016disgust}. Another previous study reported that pupil size was smaller during disgust conditioning than neutral or fear conditioning \parencite{wang2021generalization}. However, no connection between pupil size and disgust sensitivty was assessed in this study. Regarding other physiological correlates, Wang et al. (\citeyear{wang2024conceptual}) did not find an influence of interpersonal differences regarding contamination fears on skin conductance rates during their disgust conditioning experiment. 

In fear conditioning it can be said that during conditioning, pupillary size increases for the CS+ compared to the CS- \parencite{leuchs2019measuring} and is positively related to expectancy of an aversive event \parencite{koenig2018pupil}. Cognitive vigilance is also correlated to pupillary increase and higher pupil variability, while cognitive avoidance is more closely related to pupillary constriction \parencite{van2018pupil}. 

Thus, it is very difficult to formulate a hypothesis for our experiment. If participants high in disgust sensitivity perceive the CS as more disgusting and/or cognitively avoid this stimulus, this would lead to reduced pupil size. If a heightened fear of an upcoming disgust stimulus is present and they vigilantely try to learn, which stimuli predict disgusting events, they would show increased pupil size. However, regarding the fact that we know that disgust sensitive individuals show delayed extinction learning, which does not correspond to correct associative learning, we assume that pupil size is negatively correlated to disgust sensitivity during conditioning. Regarding the difference between CS+ and CS-, we hypothesize increased pupillary responses for the CS+ than the CS-.

\textbf{Attentional response during disgust conditioning}
Knowles et al. \parencite{knowles2018disgust} also investigate attentional biases as one form of cognitive bias. However, gaze control is not completely under cognitive control and can thus also be partially be interpreted as a different response than the cognitive one. A current point of view regarding the attention of disgust is that first, attention is driven towards the disgusting object due to attentional capture of aversive stimuli but is then diverted away, due to the aversive nature of the stimulus and the lack of current danger of a non - moving stimulus \parencite{armstrong2022ve}. This actually might be the reason, why disgusting stimuli are often not subject to inhibitory learning, because of a lack of attention regarding new experiences. In terms of attentional deployment during conditioning, fixation durations are consistently longer for CS+ compared to CS– during fear acquisition \parencite{michalska2017anxiety,rodriguez2023shifty,xia2021saccadic}. Longer fixation durations have been described as a correlate of stronger attentional engagement. Because we assume a lack of attentional engagement in people high in DS, we hypothesize a negative influence of ndisgust sensitivity on fixation duration during disgust conditioning.

\textbf{Our Study} 
To investigate the described variables, we applied a conceptual disgust conditioning experiment similar to Wang et al. (\citeyear{wang2024conceptual}) and using the simuli of the seminal paper by Dunsmoor et al. (\citeyear{dunsmoor2015emotional}). Additionally, we implemented variations of two further variables into the design, (1) the duration of the US (0.5 vs. 3s) and (2) the existance (9-12s) vs. non existance of an interstimulus interval (ISI) between the US and the subsequent CS. In previous studies 0.5~s \parencite{wang2024conceptual} and 3~s \parencite{schweckendiek2013learning} of US presentation led to successful conditioning and we wanted to find out, which presentation time would lead to stronger effects. The ISI was also chosen as in Wang's study whereas the lack of an ISI was implemented to assess it's influence (because without an influence, further experiments could reduce this ISI for economic reasons). 
Although our study did not specifically involve participants with diagnosed disorders, previous research suggests that non-clinical participants with high disgust sensitivity provide valuable insights into contamination-related OCD (C-OCD; \cite{abramowitz2014relevance,armstrong2017pavlovian}).

\section{Method} \label{SecMethod}
\subsection{Participants}
We ran a power analysis with the dependent variable disgust rating, since this was the dependent variable of which the fewest data was collected. We expected an increase for the CS+ from categorization to acquisition of 0.5 on the likkert scale and a decrease between acquisition to extinction of 0.2 (roughly the values in the Wang et al. (\citeyear{wang2024conceptual}) paper). Running a power analysis with 1000 simulations indicated that for an interaction between phase and condition as well as for condition to achieve a power of at least 0.9 with a level of significance of p<0.05, 42 subjects would be sufficient. To have some margin, a total of 44 participants (mean age: 22.4 years, 28 female) took part in the eye-tracking study. They were recruited through the Eyelab at the university, and all provided written informed consent. The mean disgust sensitivity score was 1.84 (SD = 0.60) for females and 1.49 (SD = 0.62) for males. These scores are lower than those reported in the original sample of the questionnaire, where the mean disgust sensitivity was 2.28 (t=-3.5, p<0.01) for females  and 1.85 (t=-2.25, p=0.04) for males \parencite{schienle2002fragebogen}.

\subsection{Materials} 
Disgust sensitivity was assessed using the German version of the Disgust Scale (FEE: Fragebogen zur Erfassung der Ekelempfindlichkeit; \cite{haidt1994individual, schienle2002fragebogen, olatunji2007disgust}). This validated questionnaire consists of 37 items spanning five distinct disgust domains: death, body secretions, hygiene, spoilage, and oral rejection. The scale exhibits excellent internal consistency overall (original $\alpha = .9$, our study $\alpha = .94$ ), with domain-specific reliabilities ranging from $\alpha = .69$ for oral rejection (our study $\alpha = .77$) to $\alpha = .85$ for death (our study $\alpha = .68$). It has been found to correlate significantly with blood-related fears ($r = .47$) and compulsiveness ($r = .25$). 

\subsection{Stimuli}
The tools and animals used as CS+ and CS- were selected from an online database (for more information see  \cite{dunsmoor2015emotional}). The same images were also used in the study by Wang et al. (\citeyear{wang2024conceptual}). These images were resized to 1000 x 1000 pixels and presented at the center of the screen. The unconditioned stimuli (US) consisted of 24 images from the DIRTI Database \parencite{haberkamp2017disgust}, including 12 images from the rotten foods category and 12 from the body products category.

 

\subsection{Experimental Procedure}
\begin{figure}
\includegraphics[width=1\textwidth]{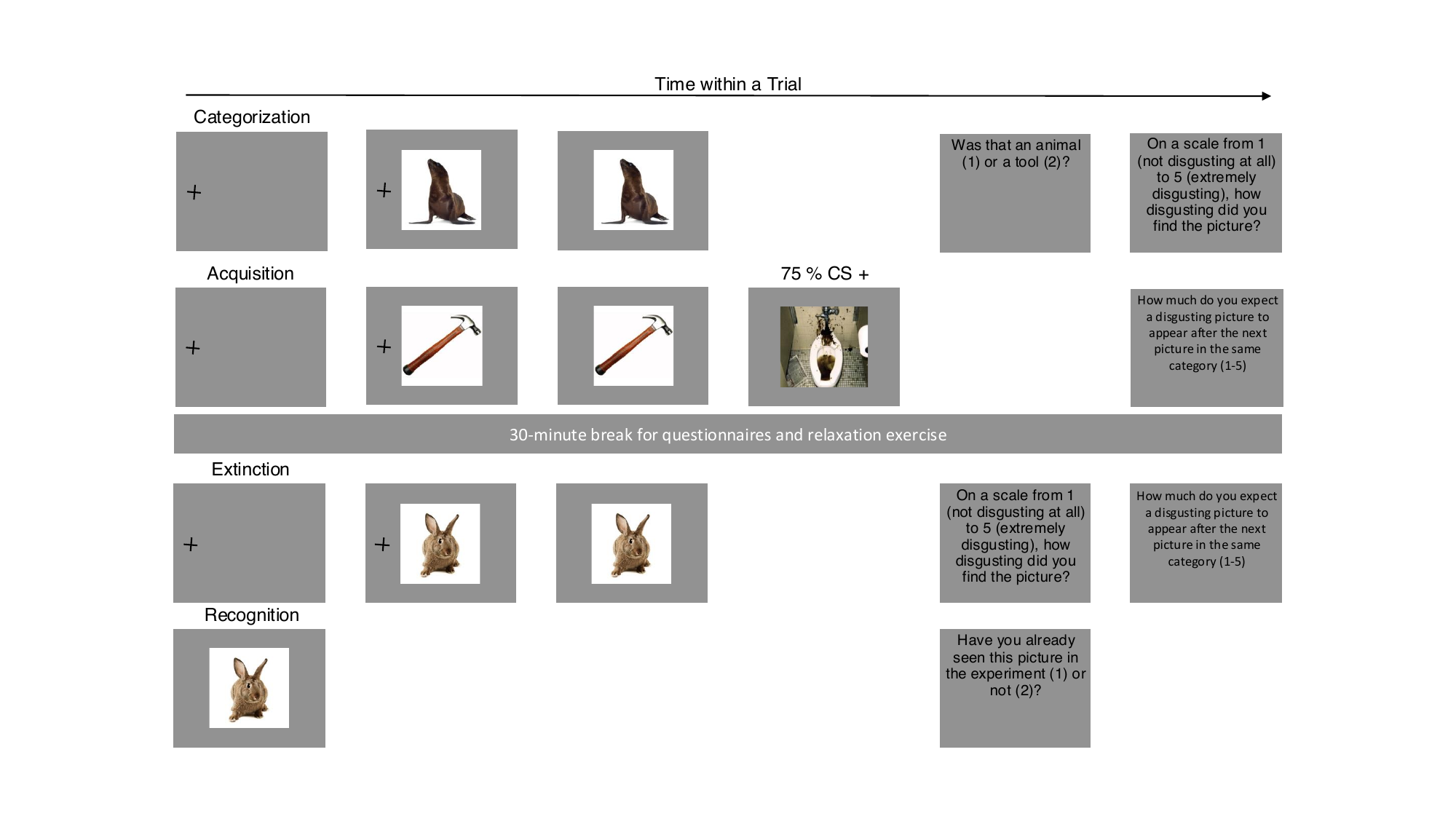}
\captionof{figure}{Experimental procedure. Horizontally, the time within a trial is represented, vertically the different experimental phases are depicted.}
\label{Procedure}
\end{figure}

Participants completed a brief questionnaire assessing their disgust reactions to rotten foods and bodily products. Their scores were used to determine the unconditioned stimulus (US) for the subsequent eye-tracking experiment, with two participants assigned to the rotten foods category and 42 participants rating bodily products as more disgusting. Before each image presentation, participants underwent a fixation check. During this check, an image was shown for 250 ms, and participants were instructed to maintain their gaze without shifting their eyes to counteract central fixation tendencies \parencite{rothkegel2017temporal}. If participants moved their eyes, the image was replaced by a mask, and the fixation check was repeated. The experiment was divided into three phases: Categorization, acquisition, and extinction. The experimental procedure is shown in Figure \ref{Procedure}.

Categorization Phase: Participants viewed 20 images (10 tools and 10 animals). After the fixation check, each image was displayed for 1.2 seconds, during which participants identified whether the image depicted an animal or a tool and rated its disgust level on a scale from 1 to 5, where 1 meant not disgusting at all and 5 extremely disgusting. 

Acquisition Phase: Participants viewed 64 new images (32 tools/animals) for 5 seconds following the fixation check. In 75\% of the trials where a conditioned stimulus (CS+) was presented, a disgust-inducing image (US) followed. The US was shown for either 0.5 or 3 seconds, with the duration counterbalanced across the CS+ image categories. As mentioned in the introduction, this manipulation aimed to assess whether the length of the US presentation affected disgust associations.
After viewing the disgust image, the first 16 participants proceeded immediately, while participants 17-44 saw a grey screen for 9-12 seconds. This variation was intended to evaluate the impact of an inter-stimulus interval (ISI) on disgust associations. 
All non-reinforced CS+ and CS- stimuli were followed by a grey rectangle. Periodically, participants rated their anticipation of seeing a disgusting image from the same category on a scale from 1 to 5. In the middle and end of the Acquisition phase, the disgust level of 6 random images (three animals, three tools), which had been seen previously, was rated. After the acquisition phase, participants completed the \textit{Fragebogen zur Erfassung der Ekelempfindlichkeit} (FEE; \cite{haidt1994individual, schienle2002fragebogen}) and participated in a 15-minute YouTube meditation session (\url{https://www.youtube.com/watch?v=4Z1RPavOX3s&t=164s}) to ensure a 30-minute break between the extinction and acquisition phases for all participants.

Extinction Phase: Participants viewed 32 new images (16 tools/animals) for 3 seconds following the fixation check. After each image, they rated its disgust level and the expectation of seeing a disgusting image from the same category. 

Recognitition Phase: In the final block, participants completed a recognition task, where they were presented with 40 images (10 tools/animals from the acquisition/extinction phases) and 20 new images (10 tools/animals). Due to a programming error, recognition data was only saved for participants 13-44. 

The experiment was programmed using MATLAB \parencite{MATLAB} and the Psychtoolbox \parencite{kleiner2007s}.

\subsection{Eye tracking data preprocessing}
Pupil data were preprocessed as outlined in König et al. (\citeyear{koenig2018pupil}). For each pupil sample, the pupil size at stimulus onset was subtracted and then divided by the pupil size at the onset. This normalization allows pupil values to represent the percentage of pupillary size relative to the size at the beginning of a trial, enabling the measurement of pupillary responses throughout the experiment regardless of the initial pupil size.

For saccade and fixation detection, we applied a velocity-based algorithm \parencite{engbert2003microsaccades, kliegl2006tracking}. An event was classified as a saccade if it met the following criteria: a minimum amplitude of 0.5 degrees of visual angle and a velocity exceeding the trial’s average by 6 median-based standard deviations for at least 6 consecutive data samples (6 ms). The interval between two successive saccades was defined as a fixation. The total number of fixations included in further analyses was 60,998.

\subsection{Statistical analysis}
To assess the significance of our results, we employed linear mixed models using the \textit{lmer} function from the \textit{lme4} package \parencite{bates2015package} within the R environment \parencite{R}. The dependent behavioral variables included reported disgust, the reported probability of the US occurring (i.e., expectancy), fixation duration, and relative pupillary size during the last second of observation. This time point was selected because prior research indicates it is when the CS type has the largest effect on pupillary size \parencite{leuchs2019measuring}.

Fixed effects in the models included stimulus type (CS+ and CS-), disgust sensitivity, and experimental phase, while the random effect was the image number. 
The inclusion of this random effect was intended to account for potential image-specific variability independent of the experimental procedure.

For each dependent variable, we initially computed a model incorporating both experimental phase and CS-type. Successive differences contrast coding \parencite{schad2020capitalize} was used for the experimental phase, allowing each phase to be compared with the subsequent one. In the next step, we computed linear mixed models for each experimental phase separately, including CS-type and disgust sensitivity as fixed effects.

To identify the best-fitting model, we first computed a model with all fixed effects and compared it to a model with one fewer fixed effect (e.g., excluding the interaction term). If the full model explained significantly more variance than the reduced model, the full model was retained; otherwise, the simpler model with fewer fixed effects that explained the most variance was chosen \parencite{bates2005fitting}.

The scripts used to compute the models, along with all experimental data are available at \url{https://osf.io/z75yq/}.

\section{Results}
\subsection{Subjective Disgust}
\begin{figure}
\includegraphics[width=1\textwidth]{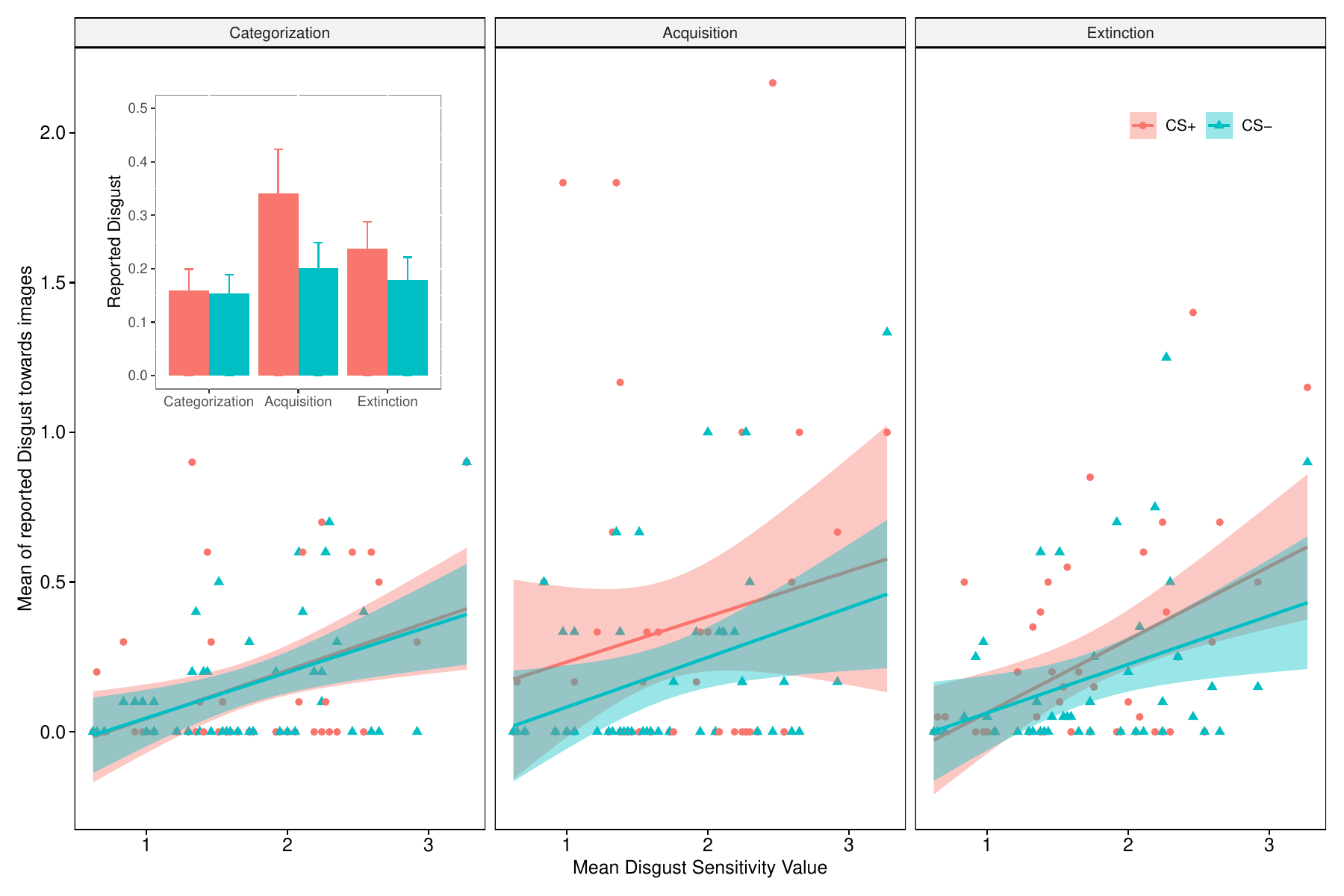}
\captionof{figure}{Reported subjective disgust for CS+ and CS- in the three experimental phases as a function of disgust sensitivity. The inserted figure shows a bar plot with errorbars representing mean and standard error of the mean.}
\label{Disgust1}
\end{figure}

The linear mixed model containing subjective disgust as the dependent variable and experimental phase and CS type as fixed effects led to the following results: There was a significant increase of reported disgust from categorization to acquisition ($ t = 3.352, p = 0.001$) and a marginally significant decrease from acquisition to extinction ($ t=-1.830, p=0.067$). 
The model revealed a main effect of CS type, such that CS+ lead to larger disgust values than CS- ($ t=-3.020, p=0.003$). Additionally, a significant interaction ($ t=-2.392, p=0.017 $) between experimental phase and CS type was found for the comparison of categorization and acquisition, such that the CS+ increased stronger from categorization to acquisition than the CS-. This interaction was not significant for the comparison of acquisition and extinction. All these results are illustrated in the inserted plot in Figure \ref{Disgust1}. 


The influence of disgust sensitivity and its interaction with the CS type on reported disgust was investigated in all three experimental phases seperately. In the categorization phase, the best fitting model removed the variable CS type completely. Since there were no unconditioned stimuli presented, CS type did not have any impact on the results.
For categorization, the model showed a highly significant influence of the disgust sensitivity, measured with the FEE, on reported disgust ($ t=6.616, p<0.0001$). In the acquisition phase, the best fitting model contained a main effect of CS type and disgust sensitivity, however no interaction between these two. This model showed a main effect of CS-type ($t=-2.818, p=0.0050$) and again a significant influence of the disgust sensitivity ($t=3.678, p=0.0003$). During extinction, again the model containing the main effect of CS-type and disgust sensitivity but no interaction created the best fit. Again, this model showed a main effect of CS-type ($t=-2.512, p=0.0121$) and again a significant influence of the disgust sensitivity ($t=9.904,  p<0.0001$). These influences are illustrated in figure \ref{Disgust1}.

Overall, it can be said that the experimental conditioning worked in a way, that CS+ increased subjetive disgust in acquisition and extinction phase stronger than the CS-. Additionally, in all experimental phases, disgust sensitivity measured with the FEE had a highly significant influence on subjectively reported disgust, confirming the validity of this instrument. Interestingly, no interaction between CS-type and disgust sensitivity was found, meaning that the relative increase in disgust for CS+ was independent of disgust sensitivity.


\subsection{Expectancy}
\begin{figure}
\includegraphics[width=1\textwidth]{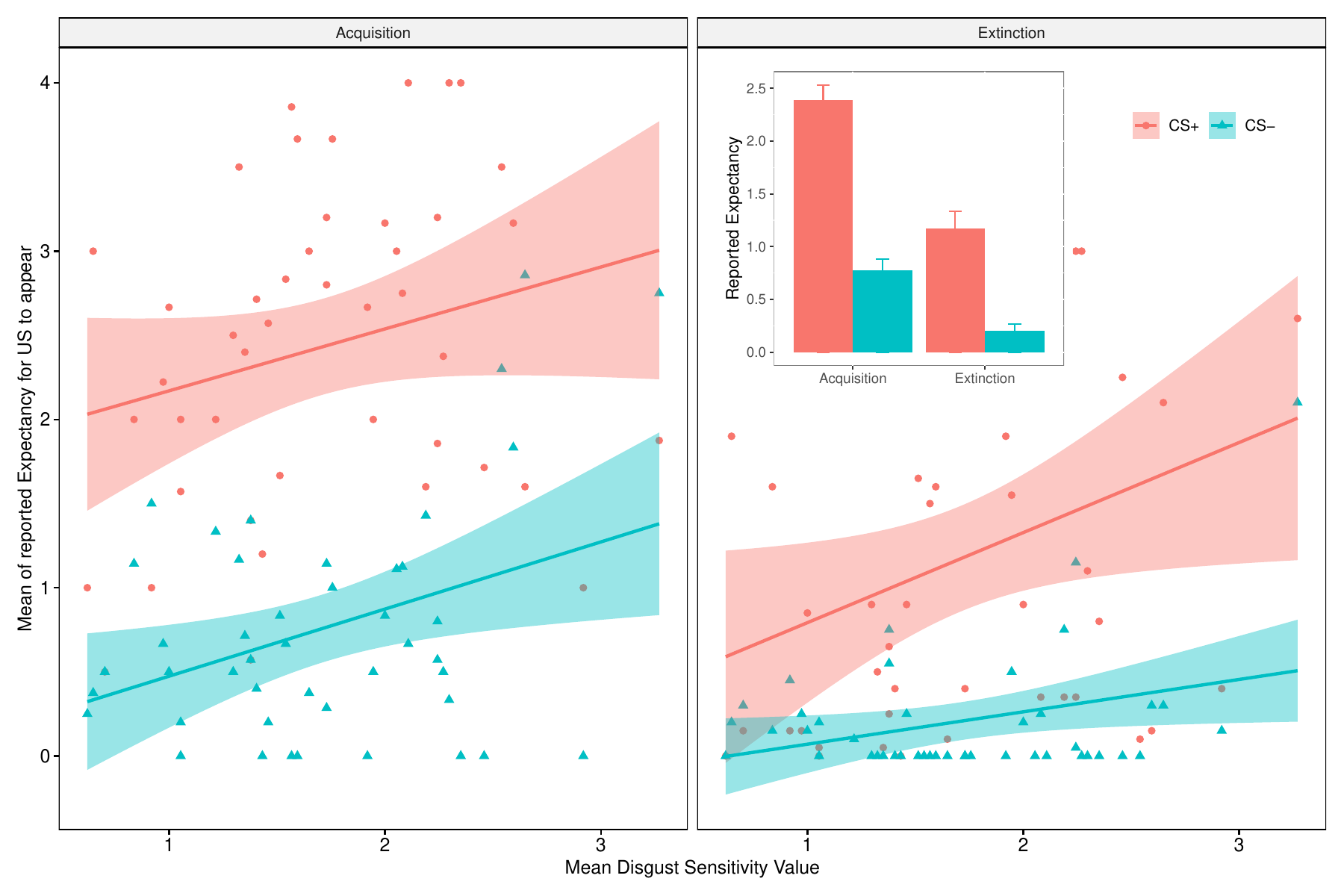}
\captionof{figure}{Reported expectancy for US to appear after CS+ and CS- in the three experimental phases as a function of disgust sensitivity. The inserted figure shows a bar plot with errorbars representing mean and standard error of the mean.}
\label{Expectancy1}
\end{figure}

The linear mixed model containing expectancy as the dependent variable and experimental phase and CS type as fixed effects lead to the following results: There was a significant decrease of US-expectancy from acquisition to extinction ($t= -15.840, p<0.0001$). No comparison for categorization is available, because expectancy was not assessed during categorization. The model revealed a main effect of CS type, such that CS+ lead to larger expectancy values than CS- ($ t=-17.313, p<0.0001$). Additionally, a significant interaction ($ t = 6.107, p<0.0001$) between experimental phase and CS type was found, such that the US-expectancy on CS+ decreased stronger from acquisition to extinction than for CS-. These results are represented in the inserted plot in Figure \ref{Expectancy1}

The influence of disgust sensitivity and its interaction with the CS-type on expectancy was investigated seperately in the two experimental phases, in which expectancy was assessed. In the acquisition phase, the best fitting model contained a main effect of CS-type and disgust sensitivity, however no interaction between these two. This model showed a significant main effect of CS-type ($t=-15.483, p<0.0001$) and a significant influence of disgust sensitivity ($t=4.376, p<0.0001$). During extinction, the full model, containing mean disgust sensitivity, CS-type and their interaction was the best fitting model. This model showed a main effect of CS-type ($ t=-2.757,  p=0.0059$), a significant influence of the disgust sensitivity ($t=10.021, p<0.0001$) and also a significant interaction ($t=-4.539, p <0.0001$). These influences are illustrated in figure \ref{Expectancy1}.

Overall, it can be said that expectancy for a US to appear after any stimulus depended strongly on disgust sensitivity in both experimental phases. The interaction effect in the extinction phase means that especially for the CS+, highly disgust sensitive individuals still expected a US to appear, which is a representation of a pathological sustained association (i.e. "I can never be sure, that it will not be disgusting again!"). 

\subsection{Pupil size}
Linear mixed models were computed for the last second of observation during each trial. As stated before, the last second of observation during conditioning is the time, when CS-type is known to produce the strongest impact \parencite{leuchs2019measuring}. The linear mixed model containing pupil as the dependent variable and experimental phase and CS-type as fixed effects lead to the following results: There was a significant increase of pupil size in the last second of observation between categorization and acquisition ($t=9.336, p<0.001$) as well as a significant decrease from acquisition to extinction ($t=-3.211, p=0.001$). However, it is important to say that the observation phase was longer in acquisition than categoriozation and extinction and pupil size on average increases during the first seconds of viewing \parencite{podladchikova2009temporal}. Thus this is an unsurprising result. The model revealed no main effect of CS type, such that CS+ did not overall lead to an increase in pupillary size. However, a significant interaction ($ t=-2.077,  p= 0.038$) of CS-type and experimental phase was found comparing categorization and acquisition phase, such that the increase in pupillary size from categorization to acquisition was larger for CS+ than CS-, agreeing with results from previous literature \parencite{leuchs2019measuring}.

The influence of disgust sensitivity and its interaction with the CS type on pupil size during the last second of observation was investigated seperately for all three experimental phases.

\begin{figure}
\includegraphics[width=1\textwidth]{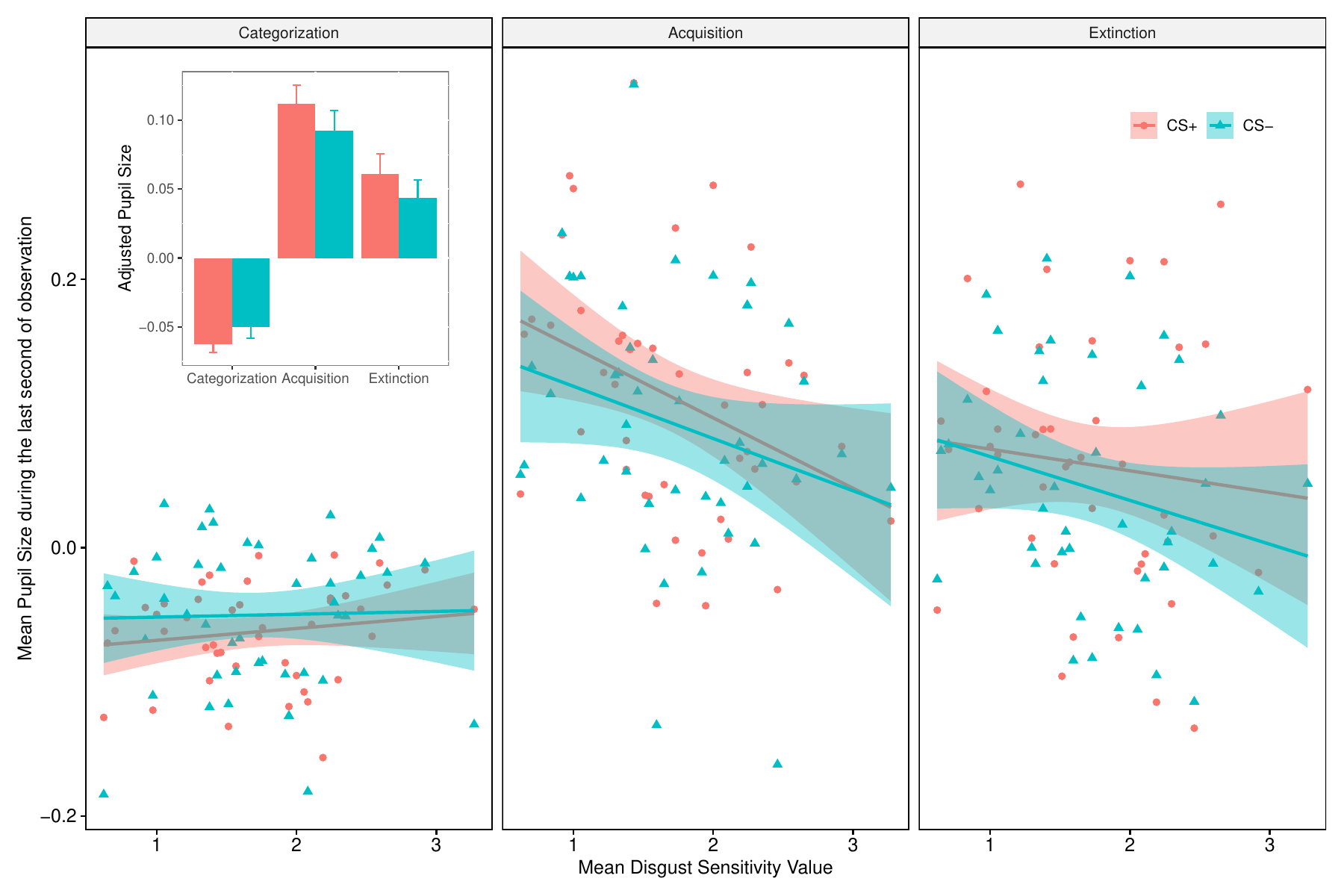}
\captionof{figure}{Mean relative pupillary size during the last second of observation for CS+ and CS- in the three experimental phases as a function of disgust sensitivity. The inserted figure shows a bar plot with errorbars representing mean and standard error of the mean}
\label{Pupil1}
\end{figure}

In the categorization phase, the best fitting model removed the variable CS type and the disgust sensitivity completely, such than there no significant influence for any of the fixed effect visible during this first experimental phase. In the acquisition phase, the best fitting model contained a main effect of CS type and disgust sensitivity, however no interaction between these two. This model showed a main effect of CS-type ($t=-2.551,  p=0.011$) and a highly significant influence of the disgust sensitivity ($ t=-7.459, p<0.001$). During extinction, again the model containing the main effect of CS-type and disgust sensitivity but no interaction created the best fit. Again, this model showed a main effect of CS-type ($t=-2.213,  p=0.027$) and again a highly significant influence of the disgust sensitivity ($t=-4.097,  p<0.001$). These influences are illustrated in figure \ref{Pupil1}.

\begin{figure}
\includegraphics[width=1\textwidth]{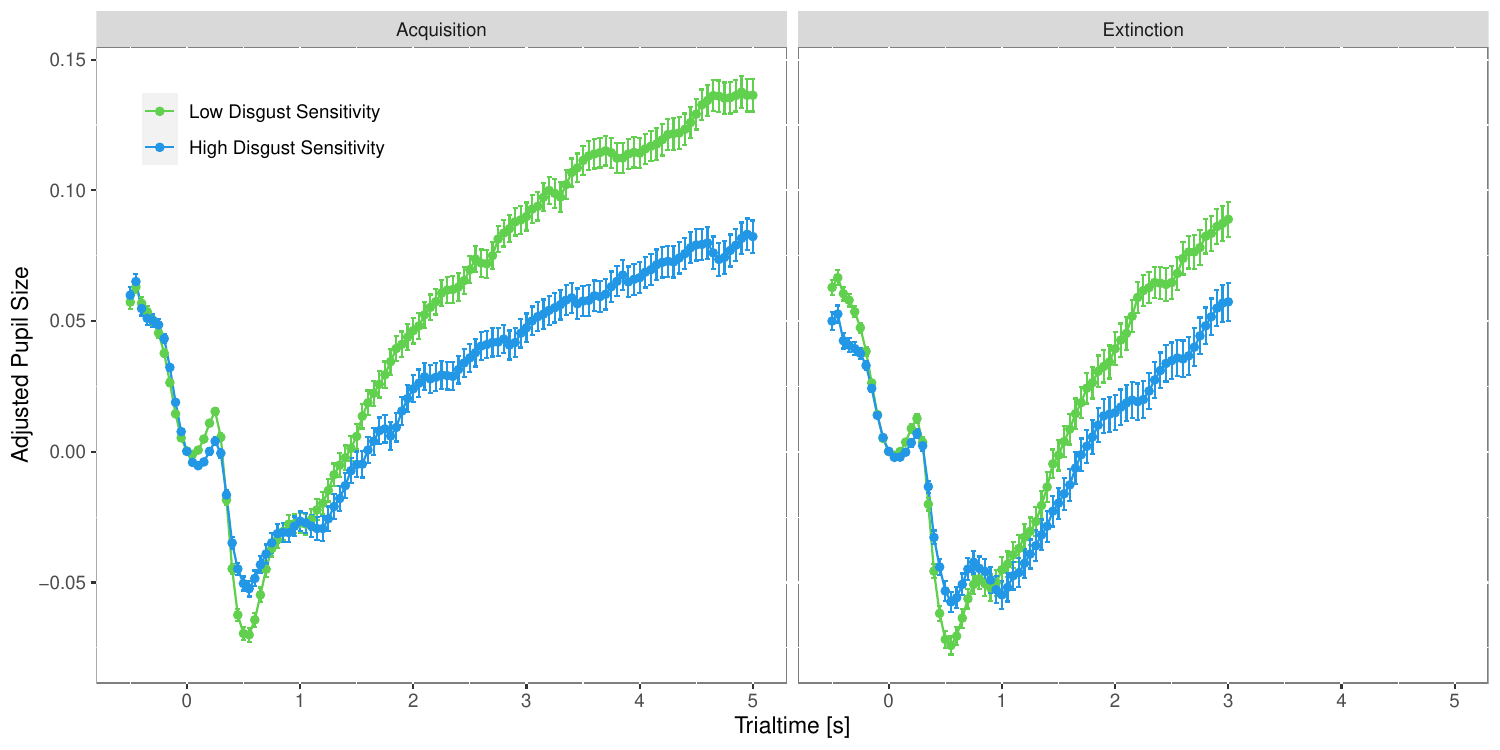}
\captionof{figure}{Pupillary development over trialtime for acquisition and extinction split by disgust sensitivity}
\label{Pupil2}
\end{figure}

As an additional figure we include the increase in pupil size as a function of trial time (aggregated for each bin of 50~ms) for acquisition and extinction split by high disgust sensitivty (>median) and low disgust sensitivty (<=median). Figure \ref{Pupil2} shows, in accordance to the results of the linear mixed models, that the pupillary increase is more pronounced for the low disgust sensitive indivuals. 

\subsection{Fixation Duration}
Linear mixed models with fixation duration as the dependent variable were computed for all fixations which landed on the CS and were below 50~ms and above 1000~ms in order to counteract strong influences of outlyers. 

The linear mixed model containing fixation duration as the dependent variable and experimental phase and CS type as fixed effects led to the following results: There was a significant increase of fixation duration between categorization and acquisition ($t=3.684; p<0.001$) as well as a significant decrease from acquisition to extinction ($t=-2.539, p=0.011$). However, as with pupil size, it is important to say that the observation phase was longer in acquisition than categoriozation and extinction and fixation durations on average increase during a free viewing trial (e.g. \cite{rothkegel2019searchers}). Thus this is an unsurprising result. The model revealed no main effect of CS type, and not interaction effect, which is not in line with previous work, where fixation durations on the CS+ were on average larger \parencite{xia2021saccadic}.

The influence of disgust sensitivity and its interaction with the CS type on fixation duration was investigated seperately for all three experimental phases.

In the categorization phase, the best fitting model removed the variable CS type completely. This is a good sign, because in this phase, CS type had no meaning yet. For categorization, the model showed a significant influence of disgust sensitivity, measured with the FEE, on fixation duration ($ t=2.107, p=0.035$), such that on average a larger disgust sensitivity led to longer fixations. 

In the acquisition phase, the best fitting model removed all fixed effects, such that there was no significant influence of disgust sensitivty, CS-type and their interaction on fixation durations. There was a marginally significant effect of disgust sensitivity on mean fixation duration ($t=-1.707, p=0.088$), however this effect was reversed compared to the categorization phase. 

During extinction, the best fitting model contained all fixed effects, CS type, disgust sensitivity and their interaction. Here, a main effect of CS-type ($t=-2.639,  p=0.008$) and a significant influence of the disgust sensitivity ($t=-2.948; p=0.003$), confirming the trend seen in acquisition, that higher disgust sensitivty leads to shorter fixations. Also, the interaction between disgust sensitivity and CS-type was significant ($t=2.875, p=0.004$). 

\begin{figure}
\includegraphics[width=1\textwidth]{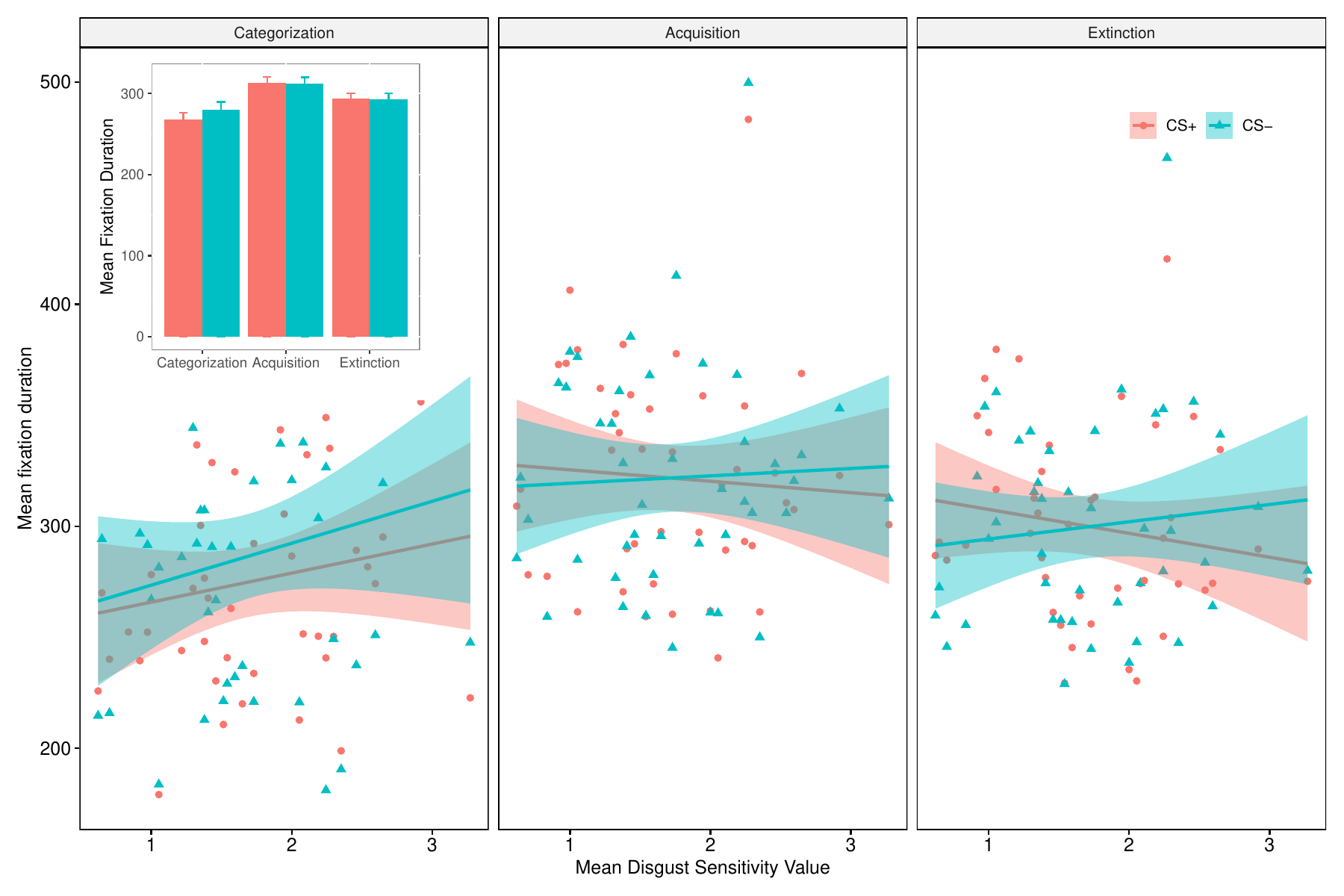}
\captionof{figure}{Mean Fixation durations for CS+ and CS- in the three experimental phases as a function of disgust sensitivity. The inserted figure shows a bar plot with errorbars representing mean and standard error of the mean}
\label{FixDur}
\end{figure}





\section{Discussion}
In a categorical disgust conditioning study, we investigated 44 participants with varying levels of disgust sensitivity. We assessed their subjective feelings of disgust and expectations of the unconditioned stimulus (US) across three phases: Categorization, acquisition, and extinction. During the experiment, we recorded fixation durations and pupil responses.

\subsection{Interpretation of the results}

The most important finding of this study, surpassing the results of Wang et al. (\citeyear{wang2024conceptual}), is the significant influence of disgust sensitivity on pupillary reactions. Disgust sensitivity was shown to significantly influence affective response, expectancy for the unconditioned stimulus (US) and pupillary responses. These findings suggest that pupillary reactions may serve as an indicator of how effectively participants learn and unlearn disgust associations, with larger pupillary reactivity during the learning process potentially leading to faster extinction learning. This aligns with the study by \parencite{leuchs2019measuring}, which states that \textit{pupil responses [...] most closely follow US expectancy.} This could imply that either heightened arousal or increased cognitive involvement during the acquisition phase — when associations are formed — as reflected by pupil dilation, may facilitate more accurate learning, which in turn is essential for effective unlearning.

The full causal relationship remains unclear, and further experiments in which pupillary size and attentional engagement during learning are experimentally manipulated could help clarify whether increased pupillary dilation reflects higher cognitive involvement, leading to better learning outcomes. Individuals with low disgust sensitivity exhibited greater pupil dilation while observing the CS images, despite rating them as less disgusting. This finding is consistent with previous research, where pupillary constriction was correlated with disgust sensitivity when participants viewed disgusting images \parencite{schienle2016disgust}. This result supports the notion that disgust may be more of a parasympathetic emotion, in contrast to fear, which is typically considered a sympathetic emotion \parencite{de2011sympathetic}.

The influence of CS-type and disgust sensitivity on fixation durations was less consistent in our study. We were unable to replicate previous findings showing longer fixation durations on the CS+ \parencite{xia2021saccadic}. One possible explanation is that our study involved conditioning at a conceptual rather than a perceptual level. To affect fixation durations, a perceptual CS+ cue may be necessary. Additionally, prior studies focused on fear conditioning, whereas our study investigated disgust conditioning, which may not influence fixation durations as strongly. Further research is needed to better understand the differing effects of fear versus disgust conditioning, as well as conceptual versus perceptual conditioning, on eye movement behavior. We did as hypothesized find a negative influence of disgust sensitivtiy on fixation duration in acquisition and extinction, which was only significant in the extinction phase. A positive influence of disgust sensitivity in the Categorization phase was also found, which was not hypothesized. A possible interpretation of these results is that during categorization, highly disgust sensitive individuals are more attentionally engaged but the presentation and expectancy of disgusting stimuli changes this pattern and turns the attentional engagement around to cognitive avoidance. Further experiments are needed to confirm this interpretation. 

While some experimental variables varied during the experiment, the detailed effects of the following independent variables are not reported in this article: the presentation time of the US (0.5 s vs. 3 s) and the inter-stimulus interval (ISI; 0 vs. 9-12 s). However, all analyses and data related to these variables are available in the OSF repository. In summary, the strongest overall increase in reported disgust was observed with a 0.5 s disgust stimulus and the ISI. Conversely, expectancy was highest when the ISI was used with a 3 s US presentation.

\subsection{Experimental limitations}

Unfortunately, a programming error led to the loss of memory data for the first 12 participants. As a result, the statistical power of the memory data is weaker than that of the other results, leading to the decision not to include it in this article. In summary, no significant difference was found between CS+ and CS-, which contradicts the findings of \parencite{dunsmoor2015emotional} and \parencite{wang2024conceptual}. However, Dunsmoor et al. used electroshocks as a presumably stronger US, and Wang et al. assessed recognition a day after the conditioning experiment, which may have heightened the salience of the conditioned stimulus. In contrast, our participants were asked to recognize images either immediately after extinction or after a 30-minute break following acquisition, which presents a clear experimental limitation.

Another experimental limitation was the variation in viewing times across different phases. While this was done for economic reasons, the unequal viewing times make it difficult to directly compare fixation durations and pupillary size between experimental phases. 

Participants were allowed to perform eye movement during the task, because we were interested in fixation durations. However, regarding pupillary reactions eye movements lead to substantial noise. Thus, recreating this experiment with only fixational eye movements and replicating the results (and additionally investigating microsaccades) would strengthen the impact of these findings. 

Last but not least, we were interested in finding which presentation time of US (0.5 to 3~s) and ISI (0 vs. 9-12~s) would lead to succesful conditioning. Thus, conditions are not completely equal. This can be interpreted as a limitation but also as a strength, because the main results apparently appear under different conditions. 

However, some differences in success of the conditioning design could be seen (see OSF) and thus, we propose the following recommendations for future eye tracking experiments using a similar design: a) a brief US presentation of 0.5-1 s is sufficient, b) an inter-stimulus interval after each US should be included to strengthen associations between CS+ and US, c) all experimental phases should have equal duration to enable meaningful comparisons of fixation durations and pupillary responses, and d) if a memory test is conducted, it should be administered the following day or later.

\section{Conclusion} \label{SecConclusion}

In a categorical Pavlovian conditioning paradigm, we demonstrated that participants with high disgust sensitivity exhibited: a) higher disgust ratings, b) greater and longer lasting expectancy for a US to appear, and c) smaller pupillary reactions during both learning and extinction. Finding b) replicates the results of Armstrong and Olatunji \citeyear{armstrong2017pavlovian}, with the addition that, in our study, the CS+ was categorical, meaning participants associated disgust with an entire category of images rather than a single distinct stimulus (see also \cite{wang2024conceptual}). Clinically, this finding is significant because it suggests that, in individuals with high disgust sensitivity, both sensory and conceptual associations related to disgust are particularly resistant to extinction. The smaller pupillary responses observed in these individuals may indicate that pupillary dilation serves as a marker of vigilance during the learning of CS-US associations. Further experiments are needed to explore this difference in pupillary response and its relationship to the anticipation of disgusting stimuli.

\section*{Acknowledgements}

We want to thank Petra Schienmann and the student assistants of the EyeLab for coordination of the lab and data collection, Nadine Möller and Nina Barkowski for their help in data collection and fruitful discussions about the project and Mathias Weymar, Christoph Szeska and Ralf Engbert for their valuable insights regarding conditioning procedures and eye tracking analysis. We also want to thank Daniel Backhaus, Lisa Schwetlick and Ursula Balk for their help in programming and conducting the pilot study for this experiment. 

\printbibliography

\end{document}